\documentclass[twocolumn,printnumbers,amsmath,amssymb,showpacs]{revtex4}
\usepackage{graphicx}

\begin{document}

\title{New jamming scenario: From marginal jamming to deep jamming}

\date{\today}

\author{Cang Zhao}
\author{Kaiwen Tian}
\author{Ning Xu$^*$}

\affiliation{Department of Physics, University of Science and Technology of China, Hefei 230026, P. R. China}

\begin{abstract}

We study properties of jammed packings of frictionless spheres over a wide range of volume fractions.  There exists a crossover volume fraction which separates deeply jammed solids from marginally jammed solids.  In deeply jammed solids, all the scalings presented in marginally jammed solids are replaced with remarkably different ones with potential independent exponents.  Correspondingly, there are structural changes in the pair distribution function associated with the crossover.  The normal modes of vibration of deeply jammed solids also exhibit some anomalies, e.g. strengthened quasi-localization and absence of Debye-like density of states at low frequencies.  Deeply jammed systems may thus be cataloged to a new class of amorphous solids.

\end{abstract}


\maketitle

Colloidal suspensions and granular materials jam into amorphous solids when they are so compact that there is no room for constituent particles to move freely.  As a simplified model to study the formation of this rigidity, a packing of frictionless soft spheres undergoes the jamming transition labeled ``J" at a critical volume fraction $\phi_c$ \cite{liu1,liu2,van_hecke,xu1,ohern}.  Point J exhibits unusual criticality in the presence of diverging length scales \cite{olsson,silbert1,wyart,drocco,ellenbroek,keys,head,xu2}, vanishing length scales in the pair distribution function $g(r)$ \cite{silbert2,donev,zhang}, and critical scalings of a variety of quantities which depend on inter-particle potential \cite{ohern,silbert1,wyart,xu2,durian}.  The normal modes of vibration in marginally jammed solids near Point J also possess some anomalous properties. For instance, low-frequency modes are quasi-localized, anharmonic, and poor in energy conduction \cite{xu2,silbert3,xu3}.  Some special vibrational features have been recently observed in experiments as well \cite{chen,ghosh,brito,kaya}.

Although great efforts have been made to understand the anomalous properties of marginally jammed solids, it has not yet been questioned whether these properties would persist when the volume fraction keeps increasing away from Point J.  In this letter, we extend the study of jammed solids to high volume fractions.  The critical scalings well-known in marginally jammed solids no longer hold in dense systems.  Surprisingly, the scalings, structure, and vibrational properties of jammed solids undergo significant changes at approximately the same crossover volume fraction $\phi_d$.  Moreover, properties of normal modes of vibration in deeply jammed solids at $\phi>\phi_d$ are beyond our understanding of both normal solids like crystals and marginally jammed solids.

The three-dimensional systems studied here consist of a $50:50$ binary mixture of $N=1000$ frictionless spheres interacting via a spring-like repulsion: $V\left(r_{ij}\right)=\epsilon \left( 1-r_{ij} / \sigma_{ij} \right)^{\alpha}/{\alpha}$ when the inter-particle separation of particles $i$ and $j$, $r_{ij}$ is less than the sum of their radii, $\sigma_{ij}=(\sigma_i + \sigma_j)/2$, and $0$ otherwise.  Periodic boundary conditions are applied in all directions.  The diameter ratio of the two species is $1.4$ to avoid crystallization \cite{note_poly}.  We set the small particle diameter $\sigma$, particle mass $m$, and $\epsilon$ to be one.  The frequency is thus in the units of $\sqrt{\epsilon / m \sigma^2}$.

It has been shown that $\phi_c$ is protocol dependent \cite{torquato,chaudhuri}.  However, critical scalings of $\phi-\phi_c$ (or equivalently pressure $p$) observed in marginally jammed solids are independent on the value of $\phi_c$ \cite{chaudhuri}.  We thus generate jammed states at desired pressures to control their distance from Point J.  We quench a random state at infinitely high temperature and at an initial volume fraction close to that at the desired pressure to the local energy minimum by applying L-BFGS energy minimization method \cite{lbfgs}.  If the pressure of the jammed state is higher (lower) than the desired value, we decrease (increase) the volume fraction by a small amount $\delta\phi$ and apply L-BFGS again.  We apply this dilation-compression process successively until a jammed packing of spheres at desired pressure $p$ is obtained.  Note that $\delta\phi$ needs to decrease once the dilation and compression switch in order to be close enough to the desired pressure.   At each pressure, we generate $1000$ distinct jammed states from independent random states at infinite temperature and take averages over them.

\begin{figure}
\includegraphics[width=0.45\textwidth]{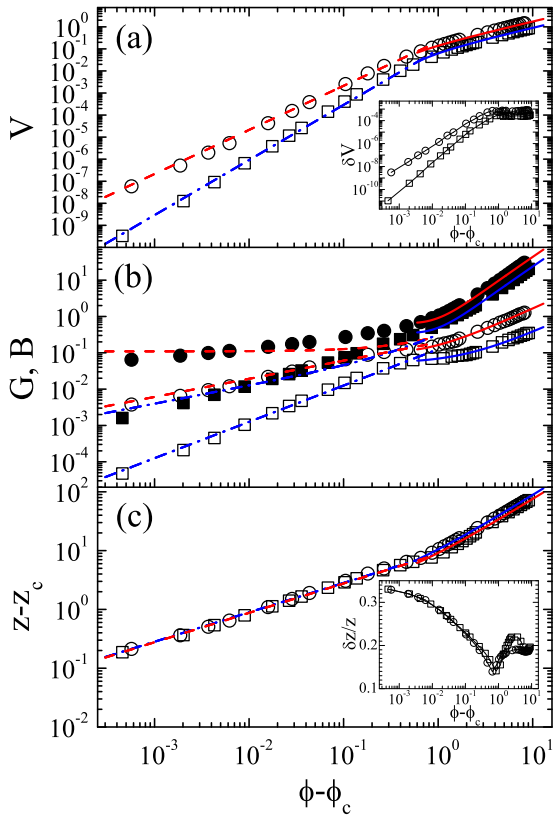}
\vspace{-0.1in}
\caption{\label{fig:fig1} (color online) (a) Potential energy per particle $V$, (b) shear modulus $G$ (empty symbols), bulk modulus $B$ (solid symbols), and (c) excess coordination number above isostaticity $z-z_c$ versus $\phi - \phi_c$ of jammed systems with harmonic ($\alpha=2$, circles) and Hertzian ($\alpha=5/2$, squares) interactions.  The red dashed (harmonic) and blue dot-dashed (Hertzian) lines are fits for marginally jammed solids according to Eq.~(\ref{marginal}), while the solid curves are fits for deeply jammed solids according to Eq.~(\ref{deep}).  The insets to (a) and (c) show the half height width of the potential energy distribution at fixed pressure, $\delta V$, and the relative spatial variation of the coordination number, $\delta z/z$ versus $\phi-\phi_c$, respectively, with the lines to guide the eye.
}
\end{figure}

Figure~\ref{fig:fig1} shows the potential energy per particle, $V$, bulk modulus, $B$, shear modulus, $G$, and coordination number, $z$ versus the volume fraction $\phi$ for both harmonic ($\alpha=2$) and Hertzian ($\alpha=5/2$) systems \cite{note}.  In marginally jammed solids ($\phi<\phi_d$), our data are fitted well with the following well-known scalings \cite{ohern}:
\begin{eqnarray}
V\sim (\phi - \phi_c)^{\alpha},~~~~~~~~~~~~~ B\sim \phi (\phi-\phi_c)^{\alpha-2}, \nonumber\\
G\sim (\phi - \phi_c)^{\alpha-3/2}, ~~~~~~~z-z_c\sim (\phi - \phi_c)^{1/2}. \label{marginal}
\end{eqnarray}
where $\phi_c\approx 0.6446$ for $\alpha=2$ and $0.6443$ for $\alpha=5/2$, and $z_c=2d$ is the isostatic value at Point J with $d$ the dimension of space.  Most of these scalings depend on the inter-particle potential, which is one of the particularities of the jamming transition at Point J.  Note that the fits to the bulk modulus using Eq.~(\ref{marginal}) slightly deviate from the data, as shown in Fig.~\ref{fig:fig1}(b).  It is because that the bulk modulus in Eq.~(\ref{marginal}) is derived from the widely accepted relation $p\sim (\phi - \phi_c)^{\alpha - 1}$, but the actual exponent is slightly larger than $\alpha -1$ \cite{xu4}.

\begin{table*}
\centering
\begin{tabular}{ccccccccccccccccc}
\hline\hline
$\alpha$ & & $V(\phi_d)$ & & $\nu_V$ & & $B(\phi_d)$ & & $\nu_B$ & & $G(\phi_d)$ & & $\nu_G$ & & $z(\phi_d)$ & & $\nu_z$
\\ [0.5ex]
\hline
2 & & $0.056\pm 0.001$ & & $1.02\pm 0.02$ & & $0.67\pm 0.06$ & & $1.71\pm 0.02$ & & $0.12\pm 0.01$ & & $1.16\pm 0.03$ & & $12.2\pm 0.1$ & & $0.97\pm 0.03$ \\
$\frac{5}{2}$ & & $0.019\pm 0.004$ & & $1.03\pm 0.02$ & & $0.36\pm 0.07$ & & $1.73\pm 0.03$ & & $0.057\pm 0.003$ & & $1.19\pm 0.05$ & & $11.4\pm 0.4$ & & $1.03\pm 0.02$ \\
\hline 
\end{tabular}
\caption{Parameters $S(\phi_d)$ and $\nu_S$ in Eq.~(\ref{deep}) for both harmonic and Hertzain systems. \label{tab1}}
\end{table*}

\begin{figure}
\vspace{-0.2in}
\includegraphics[width=0.45\textwidth]{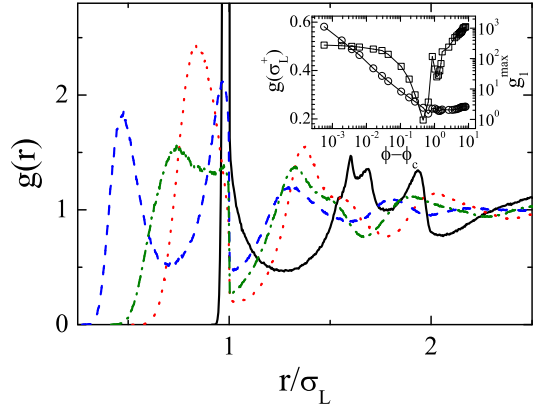}
\vspace{-2.1in}
\caption{\label{fig:fig2} (color online) Pair distribution function of large particles, $g(r)$ for harmonic systems.  The black solid, red dotted, green dot-dashed, and blue dashed lines are at $\phi=0.688$, $1.107$, $1.300$, and $2.323$.  The inset shows $g(r)$ at $r=\sigma_L^+$, $g(\sigma_L^+)$ (squares) and the first peak, $g_1^{\rm max}$ (circles) versus $\phi-\phi_c$, with the lines to guide the eye.
}
\end{figure}

The scalings described in Eq.~(\ref{marginal}) break down in the vicinity of a crossover volume fraction $\phi_d$ independent on the potentials studied here.  When $\phi>\phi_d$, Eq.~(\ref{marginal}) turns into a unified scaling with $\phi-\phi_d$:
\begin{equation}
S - S(\phi_d) \sim (\phi - \phi_d)^{\nu_S}, \label{deep}
\end{equation}
where $\phi_d\approx 1.180$ for both potentials, and $S$ denotes $V$, $B$, $G$, and $z$.  The parameters $S(\phi_d)$ and $\nu_S$ are shown in Table~\ref{tab1}.  We can tell that the exponents $\nu_S$ do not depend on the inter-particle potential.

At fixed pressure, the distribution of the potential energy of distinct jammed states is Gaussian.  The half height width of the Gaussian distribution, $\delta V$ is plotted versus $\phi - \phi_c$ in the inset to Fig.~\ref{fig:fig1}(a).  Interestingly, $\delta V$ behaves different volume fraction dependence on both sides of $\phi_d$: $\delta V\sim (\phi - \phi_c)^{\alpha}$ when $\phi < \phi_d$, and roughly a constant otherwise.

In marginally jammed solids, spatial fluctuations are significant, e.g. the distribution of local coordination number is broad \cite{silbert2,xu4}.  In the inset to Fig.~\ref{fig:fig1}(c) we plot the relative spatial fluctuation of the coordination number, $\delta z / z$ versus $\phi-\phi_c$.  When $\phi<\phi_d$, $\delta z / z$ decreases with increasing $\phi$ and drops to the minimum at $\phi_d$,  while it increases with $\phi$ when $\phi>\phi_d$.  Fluctuations in deeply jammed solids are thus not to be neglected.

Marginally jammed solids undergo some structural changes in the pair distribution function $g(r)$ during the jamming transition \cite{silbert2,donev,zhang}.  The first peak of $g(r)$, $g_1^{\rm max}$ behaves a power law divergence approaching Point J.  Meanwhile, the second peak splits into subpeaks at $r=\sqrt{3}$ and $2$ in unit of particle diameter, which become discontinuous at Point J.  Another anomalous feature is the discontinuity at $r=1$ above Point J, implying that pairs of particles that are just in contact with each other are more than those that are almost in touch \cite{silbert2}.

There are also structural signatures in $g(r)$ associated with the crossover at $\phi_d$, as shown in Fig.~\ref{fig:fig2} of $g(r)$ of large particles with harmonic interactions \cite{note2}.  When $\phi<\phi_d$, there is only one peak in $g(r)$ at $r<\sigma_L$, where $\sigma_L$ is the large particle diameter.  At $\phi_d$, a second peak starts to emerge on the left-hand side of $r=\sigma_L$, which locates at almost the same $r$ in unit of $\sigma_L$ when the volume fraction increases.  Moreover, the inset to Fig.~\ref{fig:fig2} shows that both the first peak, $g_1^{\rm max}$ and the right-hand side of $r=\sigma_L$, $g(\sigma_L^+)$ reach their minima at $\phi_d$.  These three robust changes in $g(r)$ structurally distinguish deeply and marginally jammed solids.

\begin{figure}
\vspace{-0.18in}
\includegraphics[width=0.5\textwidth]{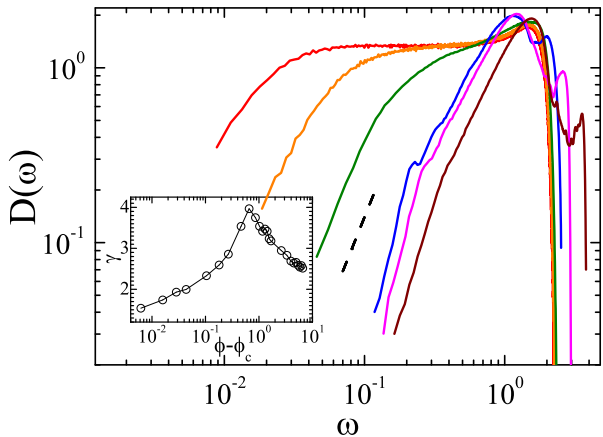}
\vspace{-2.5in}
\caption{\label{fig:fig3} (color online) Density of states, $D(\omega)$ of jammed solids with harmonic interactions.  From the left to the right, the solid curves are measured at $\phi= 0.6452$ (red), $0.6508$ (orange), $0.688$ (green), $1.107$ (blue), $2.323$ (magenta), and $5.125$ (maroon).  The dashed line shows the Debye behavior with a slope of $2$.  The inset shows the exponent $\gamma$ of the power law fit, $D(\omega)\sim \omega^{\gamma}$ to the low frequency part of $D(\omega)$ versus $\phi-\phi_c$, with the line guide to eye.  The error bars are smaller than the size of the symbols.
}
\end{figure}

\begin{figure}
\vspace{-0.2in}
\includegraphics[width=0.5\textwidth]{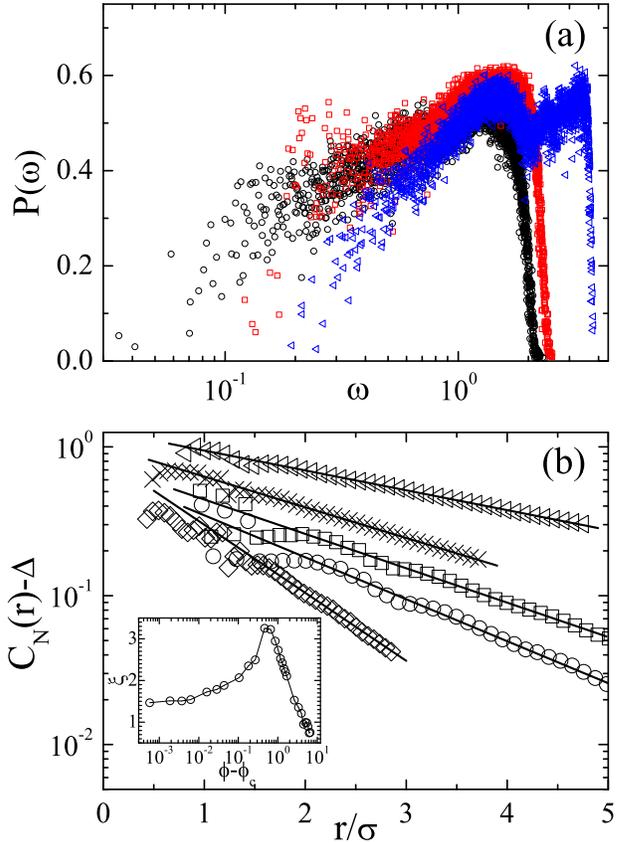}
\vspace{-0.2in}
\caption{\label{fig:fig4} (color online) (a) Participation ratio, $P(\omega)$ of jammed solids with harmonic interactions at $\phi= 0.688$ (black circles), $1.107$ (red squares), and $5.125$ (blue triangles).  (b) Normalized correlation of the polarization vectors, $C_N(r)$ of the lowest frequency mode at $\phi= 0.6508$ (circles), $0.688$ (squares), $1.107$ (left triangles), $2.323$ (crosses), and $5.125$ (diamonds).  The solid lines are fits according to $C_N(r)=C_0{\rm exp} \left( -r /\xi\right) + \Delta$.  The inset to (b) shows $\xi$ versus $\phi-\phi_c$, with the line to guide the eye.
}
\end{figure}

A special feature of marginally jammed solids is the presence of a plateau in the density of vibrational states $D(\omega)$ \cite{silbert1}.  The plateau extends to $\omega=0$ at Point J, implying the existence of excess soft modes at the jamming transition.  It has been argued that $D(\omega)$ deviates from the Debye law ($D(\omega)\sim \omega^{d-1}$) especially close to Point J \cite{silbert1}.  However, for finite size systems, low-frequency modes are sparse and discrete in frequency, leading to difficulties in determining the exact behavior of the low-frequency $D(\omega)$.  Furthermore, due to finite size effects, plane-waves hybridize with anomalous modes at low frequencies \cite{wyart,xu2}.  It is still a mystery if plane-wave-like modes could eventually dehybridize with anomalous modes and recover the Debye-like $D(\omega)$ at the low-frequency end in the thermodynamic limit.

We diagonalize the Hessian matrix using ARPACK \cite{arpack} to obtain the normal modes of vibration.  Figure~\ref{fig:fig3} shows the density of states $D(\omega)$ for both marginally and deeply jammed solids with harmonic interactions  \cite{note2}.  $D(\omega)$ at the low frequency end can be approximately fitted with $D(\omega)\sim \omega^{\gamma}$.  The inset to Fig.~\ref{fig:fig3} does not support the Debye picture, i.e. $\gamma = 2$.  Our data indicate that $\gamma<2$ close to the jamming transition at Point J.   It increases with increasing $\phi$, but does not stop at $\gamma=2$.  Interestingly, $\gamma$ reaches its maximum ($\sim 4$) at the crossover volume fraction $\phi_d$.  In deeply jammed solids, $\gamma$ decreases with increasing $\phi$.  Debye behavior is completely absent in deeply jammed solids studied here.

The anomalous density of states comes along with quasi-localization at low frequencies.  Figure~\ref{fig:fig4}(a) shows the participation ratio of normal modes \cite{note2}, defined as $P(\omega)=\frac{\left(\sum\left| \vec{e}_i^{\omega}\right|^2\right)^2}{ N\sum\left| \vec{e}_i^{\omega}\right|^4}$, where $\vec{e}_i^{\omega}$ is the polarization vector of particle $i$ in the mode at $\omega$, and the sums are over all particles.  In both marginally and deeply jammed solids, there are always quasi-localized modes with low $P(\omega)$ at low frequencies.  In marginally jammed solids, the quasi-localization is weaker with increasing $\phi$, expressed in the decrease of the number of quasi-localized modes and the increase of the minimum $P(\omega)$.  Figure~\ref{fig:fig4}(a) shows that near $\phi_d$ there are quite a few low-frequency modes with large participation ratio forming a bump in $P(\omega)$, indicating the increase of the plane-wave components.  When $\phi>\phi_d$, however, quasi-localization is strengthened with increasing $\phi$.

To quantify how the quasi-localization varies with the volume fraction, we measure the correlation of polarization vectors in the lowest frequency mode: $C(r)=\left< \vec{e}_i^{\omega_{\rm min}}\cdot \vec{e}_j^{\omega_{\rm min}}\right>$, where $\left<... \right>$ denotes the average over configurations and all pairs of particles $i$ and $j$ with a separation of $r$.  Figure~\ref{fig:fig4}(b) shows that the normalized correlation $C_N(r)=C(r)/C(0)$ can be well fitted with $C_N(r)=C_0{\rm exp} \left( -r /\xi\right) + \Delta$, where $C_0>0$ and $\Delta<0$.  The parameter $\xi$ characterizes the size of the quasi-localized regions.  The inset to Fig.~\ref{fig:fig4}(b) shows that $\xi$ is less than four particle sizes and reaches its maximum at the crossover volume fraction $\phi_d$, so the lowest frequency mode at $\phi_d$ is the least localized.

All the results shown in this letter suggest that deeply jammed solids at $\phi>\phi_d$ are new amorphous materials other than normal solids or marginally jammed solids.  It is intriguing that most properties of jammed systems undergo remarkable changes at the same crossover volume fraction.  The special features of deeply jammed solids must have significant impact on dynamics of thermal and sheared systems at very high volume fractions.  We observe that the glass transition temperature increases with the volume fraction up to $\phi_d$ and drops afterwards \cite{wang}, in consistent with a recent observation that increasing the density lowers the glass transition of ultrasoft colloids \cite{berthier2}, which is probably related to the correlation between the basin energy barrier height and quasi-localization \cite{xu3}.  We would expect similar volume fraction dependence of the yield stress.  Based on this picture, the jamming phase diagram \cite{liu1,zhang} may need corrections at high volume fractions for systems with soft potentials.  Deeply jammed solids are accessible in experiments of core-softened colloids \cite{osterman}.  We hope that the present work could open a route to explore this new kind of amorphous materials.

This work was supported by the National Natural Science Foundation of China (No. 91027001) and startup grant from USTC.


\begin{thebibliography}{50}

\item[$^*$]ningxu@ustc.edu.cn\\

\bibitem{liu1} A. J. Liu and S. R. Nagel, Nature (London) {\bf 396}, 21 (1998).

\bibitem{liu2} A. J. Liu and S. R. Nagel, Ann. Rev. of Cond. Mat. Phys. {\bf 1}, 347 (2010).

\bibitem{van_hecke} M. van Hecke, J. Phys: Condens. Matter {\bf 22}, 033101 (2010).

\bibitem{xu1} N. Xu, Front. of Phys. {\bf 6}, 109 (2011).

\bibitem{ohern} C. S. O'Hern, S. A. Langer, A. J. Liu, and S. R. Nagel, Phys. Rev. Lett. {\bf 88}, 075507 (2002); C. S. O'Hern, L. E. Silbert, A. J. Liu, and S. R. Nagel, Phys. Rev. E {\bf 68}, 011306 (2003).

\bibitem{olsson} P. Olsson and S. Teitel, Phys. Rev. Lett. {\bf 99}, 178001 (2007).

\bibitem{silbert1} L. E. Silbert, A. J. Liu, and S. R. Nagel, Phys. Rev. Lett. {\bf 95}, 098301 (2005).

\bibitem{wyart} M. Wyart, S. R. Nagel, and T. A. Witten, Europhys. Lett. {\bf 72}, 486 (2005); M. Wyart, L. E. Silbert, S. R. Nagel, and T. A. Witten, Phys. Rev. E {\bf 72}, 051306 (2005).

\bibitem{drocco} J. A. Drocco, M. B. Hastings, C. J. O. Reichhardt, and C. Reichhardt, Phys. Rev. Lett. {\bf 95}, 088001 (2005).

\bibitem{ellenbroek} W. G. Ellenbroek, E. Somfai, M. van Hecke, and W. van Saarloos, Phys. Rev. Lett. {\bf 97}, 258001 (2006).

\bibitem{keys} A. S. Keys, A. R. Abate, S. C. Glotzer, and D. J. Durian, Nat. Phys. {\bf 3}, 260 (2007).

\bibitem{head} D. A. Head, Phys. Rev. Lett. {\bf 102}, 138001 (2009).

\bibitem{xu2} N. Xu, V. Vitelli, M. Wyart, A. J. Liu, and S. R. Nagel, Phys. Rev. Lett. {\bf 102}, 038001 (2009); V. Vitelli, N. Xu, M. Wyart, A. J. Liu, and S. R. Nagel, Phys. Rev. E {\bf 81}, 021301 (2010).

\bibitem{silbert2} L. E. Silbert, A. J. Liu, and S. R. Nagel, Phys. Rev. E {\bf 73}, 041304 (2006).

\bibitem{donev}  A. Donev, S. Torquato, and F. H. Stillinger, Phys. Rev. E {\bf 71}, 011105 (2005).

\bibitem{zhang} Z. Zhang, N. Xu, D. T. N. Chen, P. Yunker, A. M. Alsayed, K. B. Aptowicz, P. Habdas, A. J. Liu, S. R. Nagel, and A. G. Yodh, Nature (London) {\bf 459}, 230 (2009).

\bibitem{durian} D. J. Durian, Phys. Rev. Lett. {\bf 75}, 4780 (1995).

\bibitem{silbert3} L. E. Silbert, A. J. Liu, and S. R. Nagel, Phys. Rev. E {\bf 79}, 021308 (2009).

\bibitem{xu3} N. Xu, V. Vitelli, A. J. Liu, and S. R. Nagel, Europhys. Lett. {\bf 90}, 56001 (2010).

\bibitem{chen} K. Chen {\it et al.}, Phys. Rev. Lett. {\bf 105}, 025501 (2010).

\bibitem{ghosh} A. Ghosh, V. K. Chikkadi, P. Schall, J. Kurchan, and D. Bonn, Phys. Rev. Lett. {\bf 104}, 248305 (2010).

\bibitem{brito} C. Brito, O. Dauchot, G. Biroli, and J. P. Bouchaud, Soft Matter {\bf 6}, 3013 (2010).

\bibitem{kaya} D. Kaya, N. L. Green, C. E. Maloney, and M. F. Islam, Science {\bf 329}, 656 (2010).

\bibitem{note_poly} Change of the particle size dispersion may affect some of the conclusions presented here, which will be studied in follow-up work.

\bibitem{torquato} S. Torquato, T. M. Truskett, and P. G. Debenedetti, Phys. Rev. Lett. {\bf 84}, 2064 (2000).

\bibitem{chaudhuri} P. Chaudhuri, L. Berthier, and S. Sastry, Phys. Rev. Lett. {\bf 104}, 165701 (2010).

\bibitem{lbfgs} http://www.ece.northwestern.edu/~nocedal/lbfgs.html.

\bibitem{note} All the results reported in this letter are valid for systems with $\alpha=3$ as well, which are not shown here.

\bibitem{xu4} N. Xu and E. S. C. Ching, Soft Matter {\bf 6}, 2944 (2010).

\bibitem{note2} We have similar observations for Hertzian systems not shown here.

\bibitem{arpack} http://www.caam.rice.edu/software/ARPACK.

\bibitem{wang} L. J. Wang and N. Xu, unpublished (2010).

\bibitem{berthier2} L. Berthier, A. J. Moreno, and G. Szamel, Phys. Rev. E {\bf 82}, 060501(R) (2010).

\bibitem{osterman} N. Osterman, D. Babic, I. Poberaj, J. Dobnikar, and P. Ziherl, Phys. Rev. Lett. {\bf 99}, 248301 (2007).

\end{thebibliography}
\end{document}